\documentclass[sigconf]{acmart}
\AtBeginDocument{%
  }

\copyrightyear{2026}
\acmYear{2026}
\setcopyright{cc}
\setcctype{by}
\acmConference[CHI '26]{Proceedings of the 2026 CHI Conference on Human Factors in Computing Systems}{April 13--17, 2026}{Barcelona, Spain}
\acmBooktitle{Proceedings of the 2026 CHI Conference on Human Factors in Computing Systems (CHI '26), April 13--17, 2026, Barcelona, Spain}
\acmPrice{}
\acmDOI{10.1145/3772318.3791531}
\acmISBN{979-8-4007-2278-3/2026/04}

\usepackage{ulem}
\usepackage{xspace}
\usepackage{tablefootnote}
\newif\ifstatus
\statustrue

\definecolor{goodgreen}{RGB}{10,143,34}
\definecolor{goodblue}{RGB}{32,0,128}

\newcommand\out[1]{}
\newcommand\rev[1]{{{#1}\xspace}}
\newcommand\moved[1]{{{#1}\xspace}}

\begin{document}


\title[Privacy and Safety Experiences and Concerns of U.S. Women Using Generative AI for Seeking SRH Information]{Privacy and Safety Experiences and Concerns of U.S. Women Using Generative AI for Seeking Sexual and Reproductive Health Information}

\author{Ina Kaleva}
\orcid{0000-0003-4303-924X}
\affiliation{
  \institution{King's College London}
  \city{London}
  \country{United Kingdom}}
\email{ina.1.kaleva@kcl.ac.uk}

\author{Xiao Zhan}
\affiliation{%
  \institution{VRAIN, Universitat Politècnica de València \\ \& University of Cambridge}
  \city{Valencia}
  \country{Spain}
  \city{Cambridge}
  \country{United Kingdom}}
\email{xzhan1@upv.es}

\author{Ruba Abu-Salma}
\affiliation{%
  \institution{King's College London}
  \city{London}
  \country{United Kingdom}}
\email{ruba.abu-salma@kcl.ac.uk}

\author{Jose Such}
\affiliation{%
  \institution{INGENIO (CSIC-Universitat Politècnica de València)}
  \city{Valencia}
  \country{Spain}}
\email{jose.such@csic.es}


\renewcommand{\shortauthors}{Kaleva et al.}

\begin{abstract}
The rapid adoption of generative AI (GenAI) chatbots has reshaped access to sexual and reproductive health (SRH) information, particularly following the overturning of \emph{Roe v. Wade}, as individuals assigned female at birth increasingly turn to online sources. However, existing research remains largely model-centered, paying limited attention to user privacy and safety. We conducted semi-structured interviews with 18 U.S.-based participants from both restrictive and non-restrictive states who had used GenAI chatbots to seek SRH information. Adoption was influenced by perceived utility, usability, credibility, accessibility, and anthropomorphism, and many participants disclosed sensitive personal SRH details. Participants identified multiple privacy risks, including excessive data collection, government surveillance, profiling, model training, and data commodification. While most participants accepted these risks in exchange for perceived utility, abortion-related queries elicited heightened safety concerns. Few participants employed protective strategies beyond minimizing disclosures or deleting data. Based on these findings, we offer design and policy recommendations—such as health-specific features and stronger moderation practices—to enhance privacy and safety in GenAI-supported SRH information seeking.
\end{abstract}

\ccsdesc[500]{Security and privacy~Social aspects of security and privacy}
\ccsdesc[500]{Security and privacy~Usability in security and privacy}
\ccsdesc[500]{Security and privacy~Privacy protections}
\ccsdesc[500]{Human-centered computing~Empirical studies in HCI}

\keywords{Women’s health, sexual and reproductive health (SRH), digital health, generative AI (GenAI), privacy, safety.}

\maketitle


\section{Introduction}
In recent years, the rapid growth of generative AI (GenAI) chatbots has transformed how people access and interact with information across diverse domains, extending beyond traditional search engines \cite{zhou2024understanding}. Built on large language models (LLMs) and trained on extensive datasets, GenAI chatbots can generate text, images, and other forms of content in response to user prompts \cite{briganti2024chatgpt}. These tools are developed by a wide range of organizations, including OpenAI’s ChatGPT, Microsoft’s Copilot, and Google’s Gemini. Notably, ChatGPT reached approximately 100 million unique users within two months of its launch, marking the fastest adoption of any online platform to date \cite{ebert2023generative}. Although initially designed for general-purpose use, GenAI tools are increasingly being applied in healthcare, offering new avenues for accessing information related to sexual and reproductive health (SRH) \cite{alhur2024redefining}.

People assigned female at birth increasingly turn to online resources to seek information on sensitive and often stigmatized SRH issues, as these platforms provide rapid and, in principle, discreet access to information \cite{flores2023internet, sellke2022unprecedented, dewan2024teen}. GenAI chatbots are particularly appealing alternatives to traditional search engines and social media due to their conversational interfaces and perceived anthropomorphism \cite{wang2023people, menon2023chatting, mills2023chatbots, zhang2024s}, ability to deliver personalized responses \cite{javaid2023chatgpt, pearson2019personalisation}, perceived anonymity \cite{mills2023chatbots}, and convenience \cite{zhang2024s}.

In 2022, the Supreme Court’s decision in Dobbs v. Jackson Women’s Health Organization overturned Roe v. Wade, eliminating the constitutional right to abortion in the United States (U.S.). Following this ruling, legal restrictions on reproductive healthcare have heightened safety risks for individuals seeking abortion and other SRH services, including surveillance, criminalization, social stigma, gender-based violence, and targeted misinformation \cite{meister2022digital, pallitto2013intimate}. Limited access to abortion care in certain states has further amplified privacy and legal concerns, as digital data could potentially be used to prosecute individuals seeking abortion-related or other SRH-related information \cite{Jones2025}.

Prior research has examined the privacy and safety behaviors of users of female-oriented technology (FemTech), particularly in the context of period-tracking apps \cite{cao2024deleted, mohan2025flowing}. The dynamic, conversational interactions between users and GenAI chatbots generate substantial amounts of personal data, often encompassing sensitive and potentially legally contentious topics, raising concerns about data breaches and misuse \cite{golda2024privacy,abu-salma2025grand}. However, research on GenAI chatbots for SRH information remains limited. Existing studies have primarily been model-centered, focusing on clinical efficacy and reliability rather than users’ experiences or the broader societal implications of these tools \cite{bachmann2024exploring, amin2024focus, wan2023chatgpt, cao2025large}.

To address this gap, we conducted semi-structured interviews with 18 participants who sought SRH information via GenAI chatbots in the post-Roe era, focusing on their privacy and safety experiences. The study aims to understand the facilitators and barriers to adopting and using GenAI chatbots in this context, participants’ perceptions of data practices, their privacy and safety concerns, and the strategies they employed to mitigate risks. Our research questions (RQs) are as follows:

\textbf{RQ1.} What factors facilitate or hinder the adoption and use of GenAI chatbots for SRH information seeking?

\textbf{RQ2.} What are users’ beliefs about data flows and practices when using GenAI chatbots for SRH information?

\textbf{RQ3.} What privacy and safety risks are users concerned about when using GenAI chatbots to seek SRH information? How do these risk perceptions differ across restrictive and non-restrictive states, and across different SRH topics?

\textbf{RQ4.} What measures or strategies do GenAI chatbot users employ (or would consider employing) to protect against these risks, and how can GenAI chatbots be improved to better safeguard users seeking SRH information?

We found that participants used GenAI chatbots to seek answers to a range of SRH queries due to the tools’ utility, usability, perceived credibility, accessibility, and anthropomorphism. Barriers to adoption included limited usefulness for serious health conditions, usability challenges, lack of perceived credibility, risk of bias, and absence of human empathy and experience (RQ1). Participants often expressed uncertainty or held inaccurate beliefs about GenAI chatbots’ data practices, including data collection, processing, sharing, and deletion (RQ2). They also perceived GenAI chatbots as posing higher privacy risks than other SRH information sources (e.g., search engines, period-tracking apps, social media, and healthcare providers), primarily due to the large volume of personal data collected and processed. Additional perceived risks included model training, government surveillance, user profiling, advertising, and insufficient regulatory protections. Participants identified several harms from potential privacy breaches of their SRH-related conversations with GenAI chatbots, such as criminalization, emotional distress, harassment, and stigmatization. Most were willing to share SRH information in exchange for utility, except for abortion-related queries and, in some cases, other stigmatized topics, such as sexually transmitted infections (STIs) or sexual orientation and gender identity (SOGI) (RQ3). Some participants employed protective measures beyond data minimization and deletion (RQ4). Based on these findings, we propose socio-technical recommendations to enhance privacy and safety in GenAI chatbots for SRH information seeking, including introducing health-specific interactive privacy features co-designed with end-users, strengthening legal and regulatory protections, implementing SRH-adapted moderation rules, and promoting greater transparency.

To our knowledge, this is the first study to examine participants’ privacy and safety experiences with GenAI chatbots in the context of seeking SRH information. While prior research has focused on technical aspects and the clinical efficacy of GenAI chatbots, our study provides a novel perspective by offering a comprehensive, in-depth understanding of user experiences, privacy perceptions, and risk mitigation strategies in the SRH context. The insights from this research can inform the development of more user-centric, privacy- and safety-preserving, and ethically designed GenAI technologies.

\section{Related Work}

\subsection{GenAI Chatbots as a Source of SRH Information}
The Internet provides a wide range of sources for SRH information, including websites, social media platforms, blogs, and forums \cite{grace2023mixed}. The emergence of GenAI chatbots is transforming healthcare and online information seeking by enhancing efficiency, accessibility, and personalization \cite{delcea2024medicine, alhur2024redefining}. For example, ChatGPT has demonstrated the potential to meet or surpass the passing threshold for the U.S. medical licensing exam \cite{gilson2024correction}. In the context of SRH, GenAI chatbots have performed well in answering questions about fertility \cite{beilby2023089, chervenak2023promise}, pregnancy \cite{khromchenko2024chatgpt}, abortion \cite{hunter2023ai}, birth control and contraception \cite{burns2024use}, endometriosis \cite{ozgor2024accuracy}, and sexual health, including STIs \cite{koh2024chatgpt, latt2025evaluation}. They have the potential to improve SRH education and literacy \cite{beilby2023089, burns2024use}, provide innovations in menstrual health management \cite{adhikary2025menstrual}, address inequalities in SRH care \cite{adhikary2025menstrual}, and optimize doctor-patient communication \cite{traylor2025beyond, zelinschi2025generative}.%

Despite their potential, GenAI chatbots have several limitations. These include hallucinations that generate unreliable content, lack of contextual understanding, inaccurate source referencing, biased training data, unclear information origins, and no interaction with professionals, all of which can pose risks to physical safety \cite{biswas2023role, beilby2024chatgpt}. For example, ChatGPT has shown inaccuracies, such as misrepresenting the safety of self-managed medication abortions \cite{mcmahon2024automating}. Performance inconsistencies have also been observed across different GenAI chatbots, including ChatGPT, Google Bard, and Microsoft Bing \cite{mediboina2024assessing, khromchenko2024chatgpt, patel2024comparative, patel2024evaluating, burns2024use}. Comparative studies indicate that Google Bard and ChatGPT-3.5 outperform Microsoft Bing and ChatGPT-4.0 in providing references and ensuring readability when addressing contraception-related queries \cite{patel2024comparative, patel2024evaluating}.

In terms of user experience, a recent study found that participants considered ChatGPT easy to use for accessing general health information \cite{al2023investigating}. Additionally, a cross-sectional study showed that people seeking health information through ChatGPT perceived it as equally or more useful than other online resources and even their doctors \cite{ayo2024characterizing}. Consequently, participants sometimes requested referrals or changed medications based on information obtained from ChatGPT. Another study examined participants’ perceptions of the accuracy of “at-home” abortion remedy information provided by ChatGPT and featured in TikTok videos \cite{sharevski2023talking}. In this study, participants were sometimes less likely to label the video content as misinformation when they knew it was generated by ChatGPT, suggesting that GenAI chatbots may be perceived as reliable sources for SRH information.

\subsection{Privacy and Safety of GenAI Chatbots for SRH Information Seeking}

With the overturn of Roe v. Wade, there has been a notable increase in research examining privacy issues associated with the digitalization of the reproductive body, particularly focusing on period-tracking apps and other FemTech-related tools and devices, which pose unique safety risks to users \cite{almeida2022bodies, malki2024exploring, mehrnezhad2023my, cao2024deleted, mcdonald2023did, dewan2024teen}. Intimate SRH data can reveal deeply personal information about a user’s lifestyle, health status, and reproductive decisions, including sexual activity, pregnancy status, abortion history, menstrual health, and fertility. Mismanagement or misuse of such data can have serious consequences, including criminalization \cite{guo2023perspectives}, workplace monitoring \cite{brown2020femtech, mehrnezhad2021caring}, discrimination \cite{crossley2005discrimination, scatterday2021no}, harassment \cite{meister2022digital}, intimate partner violence \cite{rosas2019future, pallitto2013intimate}, targeted misinformation \cite{meister2022digital}, and criminal blackmail \cite{almeida2022bodies}.

The use of GenAI chatbots to access SRH information introduces complex privacy and safety risks for several reasons. First, the large volume of sensitive data collected during open-ended, interactive, and dynamic conversations with GenAI chatbots is processed by AI algorithms, which can encourage greater engagement and often lead to increased disclosure of personal information \cite{zhang2024s, shanmugarasa2025sok}. Second, AI’s ability to infer sensitive information from user interactions or other available datasets extends the scope of data collected beyond what users may intentionally disclose \cite{alhur2024redefining, prabhod2024leveraging, staab2023beyond, li2024personal, shanmugarasa2025sok}. Consequently, seemingly neutral data can be combined to infer sensitive health conditions and personal choices; for example, discussions about menstrual cycles and symptom patterns can reveal pregnancy status or other reproductive health issues. Third, to develop and refine their models, GenAI chatbots may use training data derived from users’ conversations, potentially including personally identifiable information (PII) and other sensitive details that users did not explicitly consent to share. As a result, the model could memorize sensitive SRH information and inadvertently expose it in responses to other users’ prompts due to insufficient data protection measures \cite{carlini2022quantifying, carlini2021extracting, zhang2023counterfactual, aditya2024evaluating, shanmugarasa2025sok}.

Existing research on GenAI chatbots is primarily model-centered, focusing on technical privacy risks \cite{carlini2022quantifying, li2023multi}, and provides limited insight into participants’ privacy experiences \cite{leschanowsky2024evaluating, zhang2024s, ma2025privacy, zhan2025malicious}. A recent study investigating how users manage disclosure risks and benefits when interacting with LLM-based conversational agents found that participants frequently faced trade-offs between privacy, utility, and ease of use \cite{zhang2024s}. Inaccurate mental models often limited participants’ awareness and understanding of the privacy risks posed by these agents \cite{zhang2024s}. Moreover, human-like interactions encouraged participants to disclose more sensitive information, making it more difficult to navigate these trade-offs effectively \cite{zhang2024s}. More recent work has shown that these human-like interactions can be further exploited maliciously to prompt users to reveal even more personal data \cite{zhan2025malicious}. Another study explored privacy and security concerns among users interacting with general-purpose GenAI chatbots for mental health support \cite{kwesi2025exploring}. Through 21 semi-structured interviews, the study found that participants frequently misunderstood the protections surrounding these chatbots, conflating their human-like empathy with human accountability and assuming that regulatory safeguards such as the Health Insurance Portability and Accountability Act (HIPAA) apply. The study also introduced the concept of “intangible vulnerability,” highlighting that mental health disclosures are often undervalued by users compared to tangible data, such as financial information.%

To support user-centric design and development of GenAI chatbots for SRH information seeking, it is essential to understand participants’ experiences. In this paper, we examine the factors that facilitate and hinder the adoption of GenAI chatbots for SRH information seeking, participants’ perceptions of data flows within the GenAI ecosystem, their privacy and safety concerns, and the protection strategies they employed or would like to see implemented.

\section{Methods}
We conducted semi-structured interviews with 18 U.S.-based participants assigned female at birth who used a general-purpose GenAI chatbot to seek SRH information.

\subsection{Participant Recruitment}

Participants were eligible if they had experience using a general-purpose GenAI chatbot (e.g., ChatGPT, Gemini, Copilot) to seek SRH–related information or advice; resided in the U.S.; were between 18 and 45 years old; were assigned female at birth; and spoke English. We focused on U.S. residents aged 18 to 45 because this group is more likely to be within their reproductive span \cite{nabhan2022women}, affected by the \textit{Roe v. Wade} overturn \cite{kaplan2022overturn}, and active users of GenAI chatbots \cite{microsoft2024generative}.

We developed a screening questionnaire to collect information on participants’ GenAI chatbot use, demographics, privacy concerns, attitudes toward abortion, technical background, and the SRH topics they had searched for. The screener was hosted on Qualtrics, included a detailed study description, and obtained informed consent for data collection. Participants were recruited through Prolific. \rev{Prolific prescreening criteria included age (18--45), sex (female), and current state of residence (Alabama, Arkansas, California, Colorado, Connecticut, Delaware, Hawaii, Idaho, Illinois, Indiana, Kentucky, Louisiana, Maryland, Massachusetts, Michigan, Minnesota, Mississippi, Missouri, Montana, Nevada, New Jersey, New Mexico, New York, North Dakota, Ohio, Oklahoma, Oregon, Tennessee, Texas, Rhode Island, Vermont, Virginia, Washington, and West Virginia)\footnote{At the time of the study, abortion was legal (until viability or with no gestational limit) in California, Connecticut, Colorado, Delaware, Hawaii, Illinois, Maryland, Massachusetts, Michigan, Minnesota, Missouri, Montana, Nevada, New Jersey, New Mexico, New York, Ohio, Oregon, Rhode Island, Vermont, Virginia, and Washington, and fully banned in Alabama, Arkansas, Idaho, Indiana, Kentucky, Louisiana, Mississippi, North Dakota, Oklahoma, Tennessee, Texas, and West Virginia.}.} Screener respondents were compensated at an average rate of \$23 per hour through Prolific.

A total of 420 \out{participants} \rev{interested candidates} completed the screener, of whom 201 met the eligibility criteria. \rev{Given the exploratory, in-depth nature of our qualitative study, we did not aim for statistical representativeness \cite{denny2018qualitative}. Instead, we used purposive sampling to achieve \textit{demographic} and \textit{experiential} diversity \cite{palinkas2015purposeful}.} \rev{Specifically, we aimed to (1) include users of different GenAI chatbots to avoid overrepresenting ChatGPT users; (2) capture a range of SRH topics to enable comparisons across privacy concerns; (3) achieve proportional demographic diversity (e.g., age groups, ethnic backgrounds, gender identities); (4) balance participants with low (0--4) and high (5--10) privacy concern scores; and (5) recruit equal numbers of participants from restrictive states (where abortion is legally banned) and non-restrictive states (where abortion is legal until viability or permitted past 18 weeks) to capture diverse legal contexts.} \rev{Eligible participants were invited on a rolling basis from a pool of 210 candidates, with invitations sent in batches due to nonresponse or scheduling conflicts. In total, we extended 110 invitations and continued recruitment until we obtained a balanced sample of 18 participants and achieved data saturation within our purposively selected sample \cite{palinkas2015purposeful} (see Table~\ref{tab:demographics}).} Participants were compensated \$40 for completing the interview via Prolific. The screener instrument is provided in \S\ref{Survey}.

\subsection{Interview Procedure}
\out{We conducted~18 semi-structured interviews. The interviewer checked after every one to two interviews whether data saturation with respect to emerging insights had been reached (also further assessed during the coding process; see below), or whether additional interviews were needed. Each interview lasted between 60 and 90 minutes and was conducted via Microsoft Teams video chat.}
\moved{We conducted~18 semi-structured interviews.}
\rev{During the interviews, we took detailed notes and assessed after every one to two sessions whether data saturation had been reached; that is, whether the data continued to yield additional insights within the scope of our RQs \cite{saunders2018saturation}. This process guided the point at which no further participants were recruited, which we refer to as data saturation. The research team also assessed data saturation during the coding and theme development process, confirming that the data were sufficient to allow meaningful interpretation to answer our RQs (see below in \S\ref{DataAnalysis}) \cite{saunders2018saturation}.}
\moved{Each interview lasted between 60 and 90 minutes and was conducted via video chat using our institution’s Microsoft Teams account.}

\rev{The interview guide encompassed a broad set of questions to encourage data saturation at the individual level \cite{saunders2018saturation}.} Interviews began with warm-up questions about participants’ experiences with general-purpose GenAI chatbots for seeking SRH information and advice, including comparisons to other sources. We then explored participants’ perceptions and concerns regarding interaction quality and AI-generated output. Next, we asked about their views on data practices, including data collection, processing, use, storage, retention, sharing, and deletion. For participants without direct experience with data deletion, we asked hypothetical questions. We also investigated participants’ privacy and safety concerns, as well as any protective strategies they employed. The subsequent questions examined participants’ views on existing privacy regulations for GenAI chatbots. Finally, closing questions focused on participants’ preferences and design recommendations for enhancing privacy and safety when using GenAI chatbots for SRH information. No time limits were imposed on responses, and each interview concluded with an open prompt for additional thoughts. Our interview instrument can be found in \S\ref{InterviewScript}.

\textit{Piloting.} We conducted three pilot interviews to refine question clarity, prioritize topics, and improve question ordering. Following the pilot interviews, we removed repetitive questions, combined overlapping items (e.g., questions exploring privacy and security separately, or data storage, retention, and safeguards across different data types), and refined the wording and definitions of several items (e.g., Q13–Q17, Q43). We also added prompts about specific privacy settings available in existing GenAI chatbots (Q46). Pilot interviews were excluded from the analysis. Further minor adjustments were made as needed while conducting the 18 interviews included in the analysis, consistent with the flexibility inherent in semi-structured interviews. All interviews were transcribed verbatim, and transcripts were checked by the authors for accuracy.

\subsection{Research Ethics}
The Research Ethics Committee at King's College London reviewed and approved our study. To conduct the interviews, we used anonymous Prolific IDs and institutional Microsoft Teams links. Participants did not need a Teams account to join. All participants received a consent form and information sheet outlining study details, including example questions and potential risks. Before the interviews, participants were assured that their responses, including sensitive or potentially criminalizing information regarding their SRH, such as abortion disclosures, would remain anonymous and confidential, and that they could skip questions or withdraw without consequences. Consent for recording and transcription was explicitly obtained through both the consent form and verbally prior to the interview. \rev{Because video recording might pose risks to data confidentiality, participants were given the option to decline recording and transcription and instead allow the researcher to take notes only. No participants chose to refuse recording or transcription.} After each interview, participants had the opportunity to discuss how they felt and ask questions. They also received information about the privacy and security settings available in their most used GenAI chatbot. \rev{We also followed trauma-informed research practices \cite{hirsch2020practicing} by offering participants} mental health resources, including support from Mental Health America via the Prolific message system. No cases of distress occurred. The transcripts were fully anonymized by assigning individual IDs and removing PII, e.g., names mentioned during the interviews, or contextual identifiers, e.g., references to profession. \rev{To fully guarantee data confidentiality,} audio and video recordings were permanently deleted immediately after transcription, and all data files and transcripts were securely stored on institutional OneDrive.

\subsection{Data Analysis}
\label{DataAnalysis}
\rev{We employed a Thematic Analysis (TA) approach inspired by Braun and Clarke’s guidelines \cite{Braunvirginia2021}.} We inductively coded the transcripts using MAXQDA. Although interviews \rev{were automatically} transcribed using Microsoft Teams, we took notes of insights throughout the interview process and maintained reflexivity. Two researchers independently coded two randomly selected transcripts to develop an initial coding frame. They then discussed their coding frames and merged them into one frame after resolving any disagreements. A third researcher also provided input to refine the frame. \rev{The two researchers repeated this step, coding a different transcript each time} to iteratively expand the frame until no new codes were introduced, thereby reaching code saturation. \rev{After finalizing the coding frame,} \rev{the} two researchers re-coded all the interviews using the finalized frame. \rev{Inter-rater reliability (IRR)} ($\kappa=0.70$) \rev{was calculated using the Intercoder Agreement function in MAXQDA, with a minimum code overlap rate of 1\% at the segment level, and only for codes included in the reported themes in \rev{\S\ref{Results}}. Although the use of IRR in qualitative research is debated, as it can be seen as misaligned with the epistemological principles of TA \cite{o2020intercoder}, in this study, IRR was not used as a strict measure of reliability. Instead, it served as a tool to support reflexivity and engagement with the data \cite{o2020intercoder}. Specifically, it helped identify different interpretive perspectives, improved communicability, and clarified the application of codes (e.g., recognizing that higher-level codes were not always necessary when lower-level codes were more appropriate) to ensure coding consistency. We report our IRR value in this paper for transparency purposes.} Our final coding frame can be found at this \href{https://osf.io/mf6q3/overview?view_only=42a2e9cd1a7a48e09d8431a54f4c3aca}{[link]}.

Next, we iteratively developed \rev{and named themes and sub-themes by grouping codes to address our RQs (see \S\ref{Results}). Theme formation required deep engagement with the data, including repeated reading, note-taking, and interpretation. The first author led theme development, and the research team engaged in iterative discussions to deepen the analysis and provide additional interpretative perspectives, which were treated reflexively (embracing researchers' subjectivity) rather than perceived as disagreements \cite{olmos2023practical}. Consequently, the themes evolved significantly throughout the analysis: several were refined and new themes were developed.} \rev{For example, following initial theme development, we deepened the analysis further by examining} differences between participants from restrictive and non-restrictive states, as well as across various SRH topics, highlighting themes that were unique or shared among specific characteristics (e.g., technical background) \cite{chepp2024comparative}. \rev{Thematic refinement continued until themes became conceptually complete, and additional data generated only further examples rather than new insights, achieving thematic saturation.} We also reviewed the excerpts for each topic to further refine the themes. \rev{As part of the reflexive nature of TA, the themes reflected what the \textit{authors} saw as important for answering the RQs.}


\subsection{Author Positionality}
As researchers, we acknowledge that our backgrounds, identities, and perspectives shaped how we approached this study, interpreted the data, and engaged with participants. The authors were three women and one man \rev{with extensive experience conducting research at the intersection of human-computer interaction (HCI) and computer security/privacy, with a focus on at-risk populations}. We also recognize that residing outside the U.S., in contexts where abortion access is less restricted, might have influenced our interactions with participants living in more restrictive states. For example, our own commitments to reproductive rights could affect how we interpreted participants’ experiences and narratives; however, we mitigated this influence by maintaining reflexivity throughout the research and critically examining our assumptions at each stage.

\subsection{Limitations}
Given the private and sensitive nature of SRH information participants might have sought using GenAI chatbots, some might have been hesitant to disclose conversations involving highly personal or potentially stigmatizing content \cite{zhang2024s}. To reduce social desirability bias, we avoided leading questions and emphasized that there were no right or wrong answers \cite{grimm2010social}. This study focused exclusively on participants who used GenAI chatbots for SRH information, potentially excluding insights from individuals who chose not to use such tools; future work should examine and compare perspectives from non-users with our findings. Another limitation arose from adjustments made to the interview script after interviewing the first three participants included in the analysis (distinct from pilot participants). These three participants were not explicitly asked about their views on using GenAI chatbots for abortion-related queries, though one raised the topic unprompted, and therefore the other two might not have had the opportunity to share their perspectives. However, insights regarding abortion-related topics were obtained from the remaining 16 participants who had the opportunity to share their perspectives. For our qualitative study, the screening survey employed a single item to assess self-reported privacy concerns, with the aim of minimizing completion time. This approach provided a general indication of participants’ privacy concerns to inform participant diversification, rather than capturing multiple dimensions through a fully validated scale. Future confirmatory research could incorporate quantitative measures to test hypotheses generated qualitatively in this study. Finally, we focused on adults aged 18–45, as this group is more likely to be in their reproductive span \cite{nabhan2022women}, \rev{and to face the risk of prosecution following the \textit{overturning of Roe v. Wade} \cite{kaplan2022overturn}}. \rev{While our focus might limit insights into privacy perceptions during later life stages (e.g., menopause), four participants discussed menopause-related topics. Future research could examine privacy perspectives among older adults seeking SRH information or those not directly at risk of abortion-related prosecution.}%

\section{Results}
\label{Results}

This section reports findings from 18 semi-structured interviews conducted to address RQ1–RQ4. Participants were evenly divided between restrictive (P1–P9) and non-restrictive (P10–P18) states. Table \ref{tab:demographics} summarizes participants’ demographics, the GenAI chatbots they most frequently used to seek SRH information, the SRH topics they inquired about, and their self-reported levels of privacy concern regarding the use of GenAI chatbots (rated on a 0–10 scale), as collected in the screening survey. 
Three participants (P8, P13, and P16) self-reported having technical backgrounds. 
\out{Given the qualitative nature of this study, we use qualifiers such as “few,” “some,” “many,” and “most,” rather than numeric counts, to provide a rough indication of the prevalence of themes.} An overview and summary of all themes and sub-themes are presented in Table \ref{tab:codes-description}.

\begin{table*}[h]
\centering
\caption{\rev{Participants’ use of GenAI tools, SRH topics they sought information on, privacy concerns, attitudes toward abortion, and demographics.}}
\resizebox{0.95\textwidth}{!}{%
\begin{tabular}{l p{1cm} p{3.9cm} p{1.5cm} p{2cm} l p{1.5cm} l l p{2.5cm}}
\toprule
ID &Most used chatbot &SRH topic(s)&Privacy concern level&Attitude toward abortion as health care&State
&Status of abortion access&Gender&Age&Ethnicity\\
\midrule
P1 &Meta AI
&Menstrual bleeding; cramps; nutrition&5 (out of 10)&Disagree
&Alabama
&Banned&Woman&29&Black or African American\\
P2 &ChatGPT
&Emergency contraception; pregnancy testing&4 (out of 10)&Agree
&Texas
&Banned&Woman&30&Asian or Asian American\\
P3 &ChatGPT
&SRH symptoms (unspecified)&0 (out of 10)&Strongly Agree
&West Virginia
&Banned&Non-binary&29&White or Caucasian\\
P4 &Gemini
&Tubal ligation failure&7 (out of 10)&Neither Agree or Disagree
&Texas
&Banned&Woman&31&Black or African American\\
P5 &Microsoft Copilot
&Menstrual cycles and fertility window&2 (out of 10)&Strongly Agree
&Tennessee
&Banned&Woman&29&Black or African American\\
P6 &ChatGPT
&Menstrual cycles and nutrition&6 (out of 10)&Strongly Agree
&Texas
&Banned&Woman&28&Asian or Asian American\\
P7 &Microsoft Copilot
&Access to abortion clinics; fertility tracking for birth control&1 (out of 10)&Strongly Agree
&Indiana
&Banned&Woman&27&American Indian or Alaska Native\\
P8 &ChatGPT
&Abortion laws&7 (out of 10)&Strongly Agree
&Tennessee
&Banned&Woman&30&White or Caucasian\\
P9 &ChatGPT
&Perimenopause symptoms; STIs&4 (out of 10)&Neither Agree or Disagree
&Texas
&Banned&Woman&36&White or Caucasian\\
P10 &ChatGPT
&HPV treatment&4 (out of 10)&Strongly Agree
&Washington
&Legal&Woman&41&White or Caucasian\\
P11 &Microsoft Copilot
&Perimenopause symptoms&7 (out of 10)&Strongly Agree
&Washington
&Legal&Woman&45&Asian or Asian American\\
P12 &ChatGPT
&Side effects, costs, and access to different contraception methods&4 (out of 10)&Strongly Agree
&Washington
&Legal&Woman&22&Asian or Asian American\\
P13 &ChatGPT
&Abortion laws; LGBTQ-related resources; menstrual cramps&9 (out of 10)&Strongly Agree
&Virginia
&Legal&Non-binary&23&White or Caucasian\\
P14 &ChatGPT
&Gender-affirming hormone therapy; interpretation of hormone profile results&7 (out of 10)&Strongly Agree
&New York
&Legal&Non-binary&29&White or Caucasian\\
P15 &ChatGPT
&At-home abortion care and guidance&6 (out of 10)&Strongly Agree
&Ohio
&Legal&Woman&37&White or Caucasian\\
P16 &Gemini
&Perimenopause symptoms&3 (out of 10)&Strongly Agree
&Massachusetts
&Legal&Woman&40&Black or African American\\
P17 &ChatGPT
&Contraception; PCOS; Hepatitis B; mammography&4 (out of 10)&Agree
&California
&Legal&Woman&37&White or Caucasian\\
P18 &Gemini
&Perimenopause symptoms&9 (out of 10)&Strongly Agree
&California
&Legal&Non-binary&40&White or Caucasian\\
\bottomrule
\end{tabular}

} 
\label{tab:demographics}
\end{table*}

\begin{table*}[h]
    \centering
    \caption{Overview and descriptions of themes, sub-themes, and recommendations.}\scriptsize 
    \begin{tabular}{lp{11.5cm}}
     \toprule
\textbf{Theme}&\textbf{Description}\\
 \midrule
\textbf{Barriers and facilitators}&\\
 \midrule
Utility&Participants frequently found GenAI chatbots effective for seeking SRH information due to their personalization and level of detail. Limitations included reduced effectiveness for serious health needs, absence of “lived experiences,” and limited understanding of personal contexts, which sometimes made responses too generic.\\
Usability&Participants highlighted the ease of use, 24/7 availability, interactive interfaces, and rapid consolidation of information as key reasons for using GenAI chatbots.\\
Credibility&Perceptions of credibility were mixed. Influencing factors included source-level aspects (e.g., company reputation, perceived expertise, origin of information), content-level aspects (e.g., accuracy, timeliness, clarity), and channel-level aspects (e.g., interactivity, ability to self-correct). Participants also noted risks of inaccuracies and hallucinations.\\
Equity and accessibility&Participants appreciated the affordability of GenAI chatbots. Opinions on information equity varied: some viewed chatbots as neutral, while others expressed concerns about bias and state-specific censorship of SRH information.\\
Anthropomorphism&Some participants valued the compassionate and nonjudgmental interaction style of chatbots, whereas others perceived them as robotic.\\
 \midrule
\textbf{Beliefs about data practices}&\\
 \midrule
Data collection&Most participants believed that both conversational and non-conversational data was collected, while some thought only conversational data was collected.\\
Data processing&Participants perceived that GenAI chatbots retrieved information from databases and/or the Internet to generate responses.\\
Data deletion&Participants described various deletion mechanisms: in-app or website settings, contacting customer support, deleting their full account, closing the browser, or requesting the chatbot to delete data.\\
Data recipients and purposes of use&Participants commonly believed data was shared with the GenAI company for service improvement, with third parties for profit, and with government or law enforcement entities.\\
     \midrule
 \textbf{Perceived privacy risks}& \\
  \midrule
Excessive data collection&Participants were concerned about large volumes of contextual SRH data being collected. They felt prompted to disclose more personal information due to chatbots’ interactivity, perceived intimacy, and personalization.\\
Government access and surveillance&Participants worried that GenAI companies could be subpoenaed or surveilled by government or law enforcement, or that chatbots could flag sensitive data, potentially leading to criminalization.\\
User profiling &Participants highlighted the risk of chatbots inferring sensitive or incorrect information (e.g., demographics, health conditions, SRH choices) from interactions.\\
Model training&Participants expressed concerns about personal SRH information being used for model training and response generation.\\
Data selling and advertising&Concerns were raised regarding unwanted, offensive, or intrusive advertisements targeted based on SRH conditions.\\
Lack of regulatory protections&Participants noted a lack of confidence in existing privacy laws applicable to SRH data in GenAI chatbots.\\

 \midrule
 
\textbf{Dynamics}&\\
  \midrule
SRH topics&Most participants felt comfortable sharing SRH information, except on criminalizing or stigmatizing topics such as abortion, SOGI-related issues, and STIs. Only two participants (both with technical backgrounds) considered any SRH topic too sensitive to search in GenAI chatbots.\\
Cross-state dynamics&Concerns about seeking abortion-related information were more prevalent among participants in restrictive states, primarily due to fear of criminalization. Conversely, participants in non-restrictive states reported feeling safer, citing lower-risk demographic profiles and higher levels of trust in GenAI companies.\\
 \midrule
 
\textbf{User-driven mitigation strategies}&\\ 
 \midrule
Data minimization&Participants minimized the data provided, particularly PII, demographics, and sensitive health information.\\
Using separate accounts&Participants used separate accounts for personal and professional purposes due to privacy concerns related to conversation history.\\ 
Data management and controls&Participants engaged in data deletion and suggested control settings to reduce data collection and tracking. Exporting data copies was seen as both beneficial and potentially risky.\\ 
External privacy tools&Participants reported using, or intending to use, private browsing or privacy-preserving tools.\\
 \midrule
 
\textbf{\rev{Recommendations}}& \textbf{\rev{}} \\ 
 \midrule
\rev{Public awareness and involvement} & \rev{Educational programs to improve GenAI literacy for users and clinicians; co-design workshops as part of Participatory Action Research (PAR).} \\
\rev{Privacy by Design} & \rev{RLHF to train models to handle sensitive prompts responsibly; conservative filtering to remove or obfuscate sensitive content before user data enters the training pipeline; default opt-outs from model training; and the advancement of machine-unlearning techniques and auditing tools to ensure effective data deletion.} \\ 
\rev{Interactive privacy} & \rev{Explicit consent during sensitive conversations; gentle discouragement of oversharing; safety warnings; reminders of privacy settings; and redirection to trusted resources.} \\ 
\rev{Improved transparency} & \rev{Explainable AI (XAI) approaches including open-source models, data visualizations, alerts, and privacy nudges.} \\
\rev{GenAI regulatory protections} & \rev{Dedicated “Health” models compliant with medical privacy laws; secure deployment of GenAI chatbots by healthcare providers under strict local data controls; and co-regulation between regulatory agencies and the industry.}\\
         \bottomrule
    \end{tabular}
    \label{tab:codes-description}
\end{table*}

\subsection{Facilitators of and Barriers to the Adoption of GenAI Chatbots (RQ1)} \label{RQ1}

This section examines facilitators and barriers to adopting GenAI chatbots for seeking SRH information, including utility, usability, credibility, equity and accessibility, and anthropomorphism.

\subsubsection{Utility.} Most participants \rev{(n=15)} found GenAI chatbots useful and effective for seeking SRH information, often viewing them as more informative than other online sources or healthcare professionals: ``\textit{that was way more information than I received from my physician}'' (P5). Half of participants \rev{(n=9)} appreciated the personalization of the content: ``\textit{it makes me feel like it's more catered to what's going on with my body}'' (P5).

The majority of participants \rev{(n=10)} made decisions about their \rev{(and sometimes others')} health based on generated SRH information, including scheduling medical appointments (e.g., surgery consultations), adjusting diet and nutrition, managing symptoms (e.g., during at-home abortion care), managing conditions such as polycystic ovary syndrome (PCOS), and implementing birth control changes (e.g., contraceptive pills or fertility awareness methods). The responses also enhanced participants' understanding of SRH issues.

Almost all participants (\rev{n=17}) reported certain limitations. They preferred consulting healthcare professionals for serious or emergency concerns. Some found the \rev{AI-generated information} too generic, reflecting the chatbots’ limited understanding of personal contexts. Participants also emphasized GenAI’s lack of “lived experiences”: “\textit{It [ChatGPT] doesn’t have experience managing menstrual health. In those cases, I would probably go to social media to hear somebody’s actual experience}” (P6). Some participants also found \rev{AI-generated information} unhelpful due to distrust or a lack of novel information.

\subsubsection{Usability.}
All participants \rev{(n=18)} reported using a GenAI chatbot due to its ease of use and convenience. They emphasized several advantages of GenAI chatbots, including 24/7 availability, quick access to information, the ability to revisit past conversations, and an interactive interface. These features were seen as significant benefits compared to traditional methods, such as consulting healthcare professionals, which are often limited by time constraints: “\textit{I don't have to schedule an appointment and wait for that information. I can get it right away}” (P3). Participants also frequently highlighted the benefit of information consolidation and synthesis: “\textit{I don't have to search through different websites just to find one sentence of what I want to know}” (P5).

\subsubsection{Credibility.} Participants’ trust in the generated information varied, ranging from high to low: ``\textit{I do trust it, to be honest. I don't ever sit there and question it}'' (P9). Despite these mixed perceptions, most participants \rev{(n=14)} acknowledged the risks of SRH misinformation, hallucinations, or intentional disinformation, which could result in delayed care, harmful decisions, or even death.
To mitigate these risks, participants took a critical approach, verifying new information by consulting external sources, discussing it with peers, asking their GenAI chatbot for sources, or consulting a different GenAI chatbot. Some participants also avoided using GenAI chatbots for personal or serious health issues. A small number of participants \rev{(n=4)} did not attempt to verify the SRH information obtained from their GenAI chatbot, with one participant relying on the chatbot itself as a verification tool.

We identified several factors influencing participants' perceptions of GenAI chatbots’ credibility, categorized as \textit{source}-, \textit{content}-, and \textit{channel}-level factors, following the framework described in Ou and Ho’s \cite{ou2024factors} meta-analysis. A source-level factor was company reputation. For example, one participant trusted Google’s Gemini to provide accurate information due to the company’s reputation. However, \rev{two} participants held misconceptions about ChatGPT’s ownership, which could inadvertently affect their trust: ``\textit{I trust the Google ChatGPT a bit more when it comes to the answers I want}'' (P7). Content-level factors included perceived information quality and content fluency. Almost all participants \rev{(n=17)} considered the SRH information they received accurate, up-to-date, and readable, though some limitations were noted: ``\textit{those locations weren't able to give you an abortion because in the state of Texas, it's illegal… it's really not accurate because you weren't able to get an abortion in those places}'' (P7). While participants trusted GenAI chatbots because the content ``\textit{reads like it comes from experts}'' (P5) and cites ``\textit{reputable sources}'' (P3), some \rev{(n=5)} expressed concern that the information could be influenced by less credible sources, such as blogs, chatbot programmers, or the broader Internet, including the ``\textit{dark web}'' (P8). 
A channel-level factor, interactivity, was also identified. Some participants \rev{(n=6)} reported that their trust and overall positive attitudes were influenced by the GenAI chatbot’s ability to maintain dialogue, respond to user feedback, and self-correct: ``\textit{I have noticed that the AI will even correct itself. If they tell me something and I'm like, 'Oh, I don't think that was true,' and they were like 'Oh, I'm sorry!' They'll even regroup what they told me}'' (P1).

\subsubsection{Equity and Accessibility.} Equitable access to information was a key factor supporting adoption. First, the affordability of GenAI chatbots, noted by a few participants \rev{(n=3)}, provided a more accessible alternative to traditional healthcare services in the U.S. Second, participants highlighted access to neutral and comprehensive information relevant to individuals from diverse sociodemographic backgrounds. Some participants \rev{(n=5)} found the SRH information from GenAI chatbots to be more objective and less influenced by political, moral, or religious perspectives compared to social media, blogs, healthcare providers, and peers: ``\textit{Let’s say I'm on a YouTube video about reproductive health; sometimes you deal with people's political, […] moral, and religious views... With the AI, I do feel like they seem to be less politically charged or trying to manipulate}’’ (P1). However, others \rev{(n=6)} expressed concerns that biased responses could disproportionately affect vulnerable groups, including `\textit{people of color, women, sexual and gender minorities}'' (P8), and noted the possibility of location-specific censorship of SRH information—particularly regarding abortion and birth control. For example, P10 questioned whether ``\textit{ChatGPT is not allowed to say certain things in different states like sharing information about how to induce abortion},'' highlighting potential access barriers to abortion information due to legal restrictions.

\subsubsection{Anthropomorphism.} The majority of participants \rev{(n=10)} valued the compassionate, nonjudgmental, and private nature of GenAI chatbots when discussing sensitive SRH issues, with some even using gendered pronouns to refer to the chatbot: ``\textit{she’ll have compassion for what you say}’’ (P1). 
However, some participants \rev{(n=4)} also noted limitations, describing GenAI interactions as ``\textit{impersonal and robotic}’’ (P12); ``\textit{There wasn’t that human element of connection and being able to talk about potential fears or worries about birth control, which you could discuss with a healthcare professional}’’ (P12). 
In addition, participants expressed concerns that ``\textit{people can get triggered by those kinds of topics easily and might even have traumatic experiences that it brings up}’’ (P10) when interacting with GenAI chatbots, suggesting that these tools may not be adequately trained, like healthcare professionals, to manage the emotional risks inherent in sensitive discussions.

\subsection{Beliefs About Data Practices (RQ2)} \label{RQ2}
All participants \rev{(n=18)} reported some level of uncertainty regarding GenAI companies’ data practices, including how data was collected, processed, shared, and deleted. Participants’ beliefs about these practices are described in greater detail in the following paragraphs.

\subsubsection{Data Collection.}
Most participants \rev{(n=16)} assumed that GenAI companies collected both conversational data, such as user inputs and AI-generated outputs, and non-conversational data, including account information (e.g., login credentials, phone numbers) and technical information (e.g., IP addresses, browser types). In contrast, a small number of participants \rev{(n=2)} believed that data collection was limited solely to conversational data or user inputs\rev{, with no account or technical information being collected}.

\subsubsection{Data Processing.}
To generate output, several participants \rev{(n=8) compared GenAI chatbots to search engines and} believed that \out{GenAI chatbots} \rev{they} collected and rephrased information from the Internet \moved{or filtered data by cross-referencing keywords with online sources: “\textit{They're highlighting, ‘OK, birth control is the main thing they're looking for’ and ‘USA is the location,’ and then they cross-reference that with a whole bunch of websites online}” (P12).} Others \rev{(n=4)} believed that the systems searched \rev{their own} databases for relevant information. \rev{One participant also mentioned clicking “\textit{on the little analyzing part, and it showed like the Python code that it was using}” (P14). Another participant, with a technical background, provided a more technical explanation: “\textit{So, it definitely takes in all the data and a lot of it is encoding in a lot of different layers to sort of break down the individual responses into many different bytes. And from there, it's, sort of, able to process everything. And then, but not compare terms or verses, in a bit, to kind of give you a response back, where you take all the information, break it down into the core parts. And then use that to process a new answer through language and coding models, and then go back to you}” (P13).}

\subsubsection{Data Deletion.}
Some participants \rev{(n=6)} reported having deleted conversational data from GenAI chatbots in the past, either due to privacy concerns or simply to “\textit{organize it a little bit more}” (P6). Participants described a range of possible data deletion mechanisms, including using in-app or website settings, contacting customer support, deleting their entire account, closing browsers, or directly asking the chatbot to delete their conversations: “\textit{I would imagine that I would have to put it in a prompt to say ‘Delete,’ but then I wouldn't have really much confidence that it's done it}” (P11). Most participants \rev{(n=15)} believed that even if they deleted their data, the company would likely retain copies within internal systems.


\subsubsection{Data Recipients and Purposes of Use.}
Almost all participants \rev{(n=17)} believed that their interactions with GenAI chatbots were primarily shared with the chatbot’s parent company for various purposes, including improving services and functionalities (e.g., training AI models, maintaining personalization), performing moderation, ensuring security, and conducting research and analytics: “\textit{now it'll feed me information that's more liberal or more sexual health-oriented}” (P8). 
Participants acknowledged that their data might be shared with third-party companies \rev{(n=5)} for selling or advertising purposes, and with government or law enforcement entities \rev{(n=8)}, which are discussed in greater detail in \S\ref{RQ3}. Other potential recipients, \rev{mentioned by two participants}, included employers in cases where information was “\textit{searched through an employment-specific account}” (P8).


\subsection{Perceived Privacy and Safety Risks of GenAI Chatbots (RQ3)} \label{RQ3}
\subsubsection{Prompts and Disclosure Practices.}
Participants used GenAI chatbots to seek information related to menstrual health and fertility, perimenopause, sexual health, pregnancy, contraception methods, abortion, hormonal health, and breast health (details can be found in Table \ref{tab:demographics}). While some participants \rev{(n=7)} asked general, non-personalized questions (e.g., “\textit{Can HPV ever be cured?}” — P10), \textbf{\rev{many} \rev{(n=11)} disclosed personal information}, such as demographics (e.g., age, gender, ethnicity, location), physical characteristics (e.g., weight), or specific SRH details. For instance, participants shared details about their menstrual health, including cycle start and end dates to identify fertile days, menstrual irregularities, conditions such as PCOS, and perimenopause symptoms: “\textit{I was like ‘Here are my symptoms. Could these be perimenopause?’}” (P11).

Additionally, two participants sought tailored abortion-related information. One participant, residing in a restrictive state, searched for nearby abortion clinics to support a friend experiencing pregnancy scares, disclosing their current location. The other participant, located in a non-restrictive state, asked Gemini for guidance during an at-home abortion, disclosing specific symptoms they were experiencing: “\textit{I would message on Gemini ‘When should I be concerned about the amount of blood that I have and seeing with a medical at-home abortion?’ […] But I'm very specific with the symptoms that I'm having, I'll just ask straight out like ‘These are the symptoms, what could be the issue?’}” (P15). Another participant disclosed their intent to undergo gender-affirming hormone therapy: “\textit{I was like, if I'm doing this hormone therapy, am I gonna have to get surgery?}” (P14). They also shared their hormone blood test results to seek interpretation: “\textit{I was like, ‘Hey, these are my results of my tests. What could this mean?’}” (P14).

\subsubsection{Privacy Risks of Using GenAI Chatbots for Seeking SRH Information.}
Many participants \rev{(n=11)} felt “\textit{vulnerable}” (P14) and considered GenAI chatbots riskier than other SRH information sources they had used, including search engines, social media, period-tracking apps, and healthcare professionals: “\textit{I don't have a lot of personal conversations with my general search engine}” (P1). Participants identified several privacy risks, including excessive data collection, government access and surveillance, data selling and advertising, model training and memorization, and, more generally, personal identification, lack of transparency, and hacking. The following subsections focus on the privacy risks that participants perceived as most prominent or unique to GenAI chatbots.

\paragraph{Excessive data collection.}
Many participants \rev{(n=10)} expressed concerns about the collection of large volumes of contextual and intimate \textit{SRH} data in a conversational format: “\textit{ChatGPT is like a whole another level of that, where you can literally dump your subconscious. People I know are like, ‘Hi, put my entire blood, my entire body scan, medical records on ChatGPT’ (...) People really use it in a very intimate way}” (P14). We identified three factors contributing to excessive data disclosure and collection. First, participants felt influenced to disclose more information due to the interactive nature of GenAI chatbots, which, unlike other tools, proactively encouraged self-disclosure through follow-up questions: “\textit{ChatGPT is the only place where it would proactively ask me for information}” (P6). Second, participants described GenAI chatbots as providing an “\textit{intimate}” (P1) space that increased comfort with sharing details: “\textit{I do feel like the fact that it feels more intimate, you're gonna be more comfortable giving more details}” (P1). Third, participants were inclined to trade additional personal information for greater personalization: “\textit{For some people, if they want more specific stuff, they might spit out information like their city or age}” (P12).

\paragraph{Government access and surveillance.}
Several participants \rev{(n=6)} recognized that government bodies could potentially access chatbot conversation data via subpoenas for legal enforcement. Although some acknowledged that government access could be useful in other criminal contexts, its implications for reproductive rights made it a “\textit{double-edged sword}” (P13). Participants often compared this risk to privacy controversies surrounding period-tracking apps or scenarios “\textit{when law enforcement seeks search engine history of people they're investigating criminally}” (P18), suggesting that these perceptions might be shaped by publicized legal cases involving other digital tools. Nevertheless, participants expressed uncertainty about how GenAI companies would handle legal compliance due to the absence of such events to date: “\textit{We haven't really seen a case like that yet, where the government's challenged them and saw how they responded} [...] \textit{I don't know if ChatGPT stands on giving up the information willingly}” (P13).

Furthermore, many participants \rev{(n=9)} expressed concerns about government access to and surveillance of sensitive SRH-related conversations, including abortion, to “\textit{ensure they [people] are obeying the laws}” (P7). Specifically, they worried that conversations could be flagged as criminal based on users' geographic locations or sensitive keywords: “\textit{I feel like certain prompts are like maybe red flags... and using the IP address or something, they can say ‘Hey, this person is in Texas and they’re looking up this information’}” (P4). One participant noted, “\textit{AI knows that it’s [abortion] not legal here}” (P9), indicating that participants perceived GenAI chatbots as capable of recognizing illegal activities. In contrast, some believed that abortion-related conversations would either not be treated differently or would be managed with increased caution under “\textit{special guidelines for more sensitive topics}” (P13). While most participants \rev{(n=12)} focused on legal concerns related to abortion, a few \rev{(n=2)} also raised concerns about legal issues involving SOGI-related conversations in states with limited protections for LGBTQ+ rights.

\paragraph{User profiling.}
When prompted, all participants \rev{(n=18)} stated that GenAI chatbots inferred details about their health status, lifestyles, sexual activity (e.g., number of sexual partners), interests, demographics, sexual orientation, locations, legal jurisdictions, political views, or emotional states based on their inputs: “\textit{It makes up a little bit of a picture about who I am and what’s concerning me}” (P11). For instance, P7 noted that the GenAI chatbot could infer “\textit{that you’re probably pregnant or you’re not pregnant}.” Another participant highlighted that GenAI chatbots could infer reproductive choices, observing that they were “\textit{someone that was not interested in having any more children and someone that was looking for a solution to that}” (P4). Most participants \rev{(n=13)} reported feeling comfortable with these inferences.

However, some participants \rev{(n=6)} recognized and felt uncomfortable with the risks posed by inaccurate inferences about their reproductive health or choices, which could lead to potential harm: “\textit{I might be curious about what an abortion is like or how to access it [...] And then if I happen to get pregnant later on, and then I already had that in my history, I could see how it would be misinterpreted. Like, if I just had a miscarriage and they're like 'Oh, no, you definitely had an abortion!'}” (P10). One participant noted that the risk of inaccurate inferences was higher in GenAI chatbots than in period-tracking apps due to the broader range of SRH data collected. Another participant described using a shared account with their partner, which resulted in the chatbot revealing the partner’s health information. Although the participant found it “\textit{funny}” (P14), using a joint account raised concerns about collective data potentially generating inaccurate inferences about the account holder or, as in this case, accidentally revealing another user's data.

\paragraph{Model training and memorization.}
Some participants \rev{(n=4)} expressed concerns about the use of personal SRH data for training purposes and generating new responses. These concerns, while not always explicitly stated, might reflect perceived risks of model memorization: “\textit{I think that the data from Google Gemini is saved and used to generate other results. Whereas for healthcare providers, they're not using your information to provide additional resources to others}” (P16). Participants also noted that GenAI companies derived financial benefits from model training.


\paragraph{Data selling and advertising.}
Many participants \rev{(n=12)} expressed concerns about the sale of data to third-party companies, such as social media platforms (e.g., TikTok, Meta), marketing firms, data brokers, or even the dark web. They feared this could lead to intrusive and inappropriate advertisements related to their health conditions: “\textit{If you search [...] that you were having pain in your private area and then they show you an ad about getting STD [sexually transmitted disease] \textit{tested, that could be offensive}”} (P1). Participants also described receiving ads tied to their conversations as “\textit{weird, creepy, and freak me out a little bit}” (P6). Similar to concerns about government access, worries regarding data selling and advertising were shaped by “\textit{scandals in the past}” (P13) involving other digital platforms.

\paragraph{Lack of regulatory protections.}
Most participants \rev{(n=13)} were unaware of, or lacked confidence in, the privacy regulations governing GenAI chatbots: “\textit{ChatGPT is not bound to any HIPAA laws or any health privacy laws. So, you have to assume that those data are just now out for consumption. [...] There is no confidentiality or privacy}” (P8). Nearly all participants expressed a desire for stronger regulatory protections, such as extending medical data privacy laws to GenAI chatbots (similar to “\textit{AI nurses}” in healthcare systems, P8), enhanced safeguards for minors, and stricter legal consequences for data breaches.


\subsubsection{Privacy and Safety Concerns Across Different SRH Topics and Data Types.}
Most participants \rev{(n=15)} felt comfortable or neutral about using GenAI tools for seeking SRH information. This was largely because they did not perceive SRH topics as particularly sensitive or risky: “\textit{If the data is at risk for being shared, it's just a normal human being asking about sexual and reproductive health}” (P2). Other reasons included trust in the GenAI company, not disclosing personal information, not perceiving risks as personally relevant (e.g., being beyond reproductive age, living in a non-restrictive state), employing protective measures, and a broader sense of resignation toward data privacy risks. Additionally, one participant expressed altruistic attitudes, likening data usage for AI model training to a research assistant taking notes in a doctor’s office.

These views often, but not always, shifted for \textbf{abortion} topics \rev{(n=12)}, and in some cases, \textbf{SOGI-related disclosures} \rev{(n=2)}, due to fear of criminalization: “\textit{I believe it's fairly benign again because I'm primarily researching perimenopause. I understand it would be much more sensitive if I was seeking information on abortion care and especially abortion care if I live in a state where that was criminalized}” (P18). Some participants \rev{(n=3)} also expressed concerns about targeted harassment from individuals opposed to SRH choices: “\textit{Pro-life people or anyone on the Internet could start harassing these people and sending death threats}” (P12).

\rev{Beyond criminalization, participants (n=8) raised concerns about stigma} \rev{and emotional distress associated with certain SRH topics}. \moved{For instance, some \rev{(n=3)} briefly mentioned \textbf{STIs} as a sensitive SRH topic due to its associated stigma: “\textit{damaging to their personal image, to their confidence}” (P13). Another participant noted that exposure of sensitive SRH conditions, such as \textbf{endometriosis}, could lead to being “\textit{seen as someone who is disabled in some way or not meeting society's standards for what a healthy person looks like}” (P8).}

\moved{Additionally, a few participants \rev{(n=2)} acknowledged the risk of negative professional consequences as a result of exposing general SRH information: “\textit{It wouldn't look good to an employer to have your sexual and reproductive information out in the open}” (P17).} \moved{Only two participants, both with technical backgrounds, considered \textbf{any type of SRH information} too sensitive to seek via GenAI chatbots. They were also the only ones who described more complex risk mitigation strategies (e.g., using a VPN and GenAI privacy settings) or extreme strategies (e.g., avoiding GenAI entirely), even when residing in non-restrictive states.}

\subsubsection{Perspectives Across Restrictive vs. Non-restrictive States.} \label{cross-state}
Participants frequently highlighted cross-state dynamics, even without prompting. Among participants living in non-restrictive states, many \rev{(n=5)} indicated they used or would use GenAI chatbots for seeking abortion-related information. These participants felt safe because of their geographic location or demographic background, which placed them at lower risk of prosecution related to abortion, and/or because of their trust in the GenAI company. The remaining participants \rev{(n=4)} stated they would not use GenAI chatbots for abortion-related queries, preferring more reliable sources or fearing legal consequences, even when residing in more liberal states: “\textit{As someone who is American and how we've seen lots of potential subpoenas in the past for personal records in regard to period-tracking apps, in regard to people seeking medical assistance, even though I'm not at risk of anything currently, it's still something where I'd rather not have my data be available and know it could be subpoenaed for any reason like that}” (P13). \rev{Some of these participants believed that GenAI chatbots would indirectly “refuse” to answer abortion-related questions, or that their ability to respond to abortion-related questions is limited, even within jurisdictions where abortion is legal: “\textit{Because of the legality around it, it'll usually say like ‘seek professional healthcare or see a healthcare provider’}” (P15).}

In contrast, only a small number of participants \rev{(n=2)} residing in restrictive states indicated that they would use or had used a GenAI chatbot to seek abortion-related information. The majority \rev{(n=6)} cited legal risks in abortion-restrictive states as their primary reason for avoiding GenAI chatbots: “\textit{If you’re in a state in the United States where abortions are effectively outlawed at this point, you should assume that those data are going straight to the authorities and that you can be arrested, fined, charged, or whatever you want to call it, for searching that}” (P8). \rev{Aside from differences in GenAI use and abortion-related queries, we did not identify clear differences in other protection strategies, reported in detail in \S\ref{RQ4}, between participants in restrictive versus non-restrictive states based on the qualitative data.}

\subsection{Risk Mitigation Strategies (RQ4)} \label{RQ4}

Participants described only a few privacy risk-mitigation strategies that they had used previously or would consider using, which are detailed in the following subsections. Reasons for not employing such strategies included lack of concern, limited awareness of available options, or trust in the GenAI company.

\subsubsection{Behavioral Strategies}

\paragraph{User-driven data minimization.}
Most commonly, participants \rev{(n=15)} intentionally withheld information during their interactions with GenAI chatbots, including PII (e.g., names), details about their SRH, and demographic information. As one participant explained, “\textit{I wanted to ask Copilot about miscarriage, but I didn’t because I didn’t want to share that}” (P5). Many participants re-framed their inquiries as more general questions or avoided using GenAI chatbots altogether, “\textit{I don't want to be very specific like, `Oh, I'm a 20-year-old woman who lives in this place and my cycle is X amount of days long' [...] I don't want it to have that information at all because I'm afraid of it}” (P6). Another participant described their reluctance to exchange personal information for more personalized responses, “\textit{I’m not giving it super specific information, I’m not receiving specific information back. So, I’m getting general information in return for the general information that I’m putting in}” (P6). These accounts illustrate participants’ willingness to protect their privacy, even when doing so reduced the utility of the chatbots.

\paragraph{Using separate accounts.}
One participant reported creating a separate account specifically to seek information about period cramps: “\textit{I am very wary, and the only time I've ever done it, I made a separate account that didn't have anything attached to it, and I use an Incognito browser}” (P13). Another participant maintained separate accounts for personal and professional use due to privacy concerns.

\subsubsection{Technical Strategies}

\paragraph{Data management controls.}
Only a few participants \rev{(n=3)} actively engaged with the privacy and security settings of their GenAI chatbots. These actions included turning off memorization or personalization features, opting out of data use for training purposes, enabling temporary chat or auto-deletion settings, and setting up multi-factor authentication. \rev{Two} participants also reported managing “Cookie” permissions. \rev{Many} participants \rev{(n=14)} deleted or would delete their conversational data in the future due to feelings of embarrassment or privacy concerns, and some explicitly expressed a desire for increased transparency and greater user control over data retention. Additionally, one participant suggested that users should have the option to export sensitive data, such as abortion-related conversations, and then have that data immediately deleted. Participants proposed several improvements to data management, including granular opt-out settings that would allow users to control data collection and sharing, the ability to manage inferences made by the GenAI chatbot, and visible data collection disclaimers prior to submitting sensitive queries.

\paragraph{External privacy tools.}
Some participants \rev{(n=4)} reported using or considering the use of incognito browsing modes: “\textit{I did literally everything. I was on VPN [Virtual Private Networks], on Incognito browser all that. So, they were like OK, this random user is having period cramps}” (P13). Additionally, \rev{some} participants \rev{(n=3)} recommended integrating private browsing modes directly into GenAI chatbots to ensure that user data was not tracked or saved: “\textit{So that I know it's not tracking and it's not collecting and saving this data}” (P6). One participant mentioned the potential use of privacy-preserving browsers such as Tor and DuckDuckGo, though they noted, “\textit{in ChatGPT I think that you can't use it unless you have your home IP. Like, I can tell if you have an IP blocker on}” (P8). Similarly, the use of VPNs was considered a potential approach by \rev{a few} participants \rev{(n=5)}, although \rev{some} did not elaborate on specific experiences using VPNs in the context of GenAI.

\section{Discussion}
In this section, we present our key findings regarding participants' experiences using GenAI chatbots for seeking SRH information. After summarizing the main takeaways, we highlight points for consideration grounded in these findings and propose actionable design and policy recommendations.

\rev{\subsection{From 24/7 Personalized Support to Censorship and Bias: Key Factors Shaping GenAI Chatbot Adoption (RQ1)}}
\rev{Our findings show that participants actively used, trusted, and sometimes preferred GenAI chatbots over healthcare professionals, valuing their perceived effectiveness, constant availability, accessibility, personalization, and compassion. Participants frequently made health-related decisions based solely on information provided by these chatbots. Although GenAI chatbots could enhance patient education and support clinical decisions \cite{maity2025large}, they should be considered a supplementary tool and not a replacement for professional medical consultation.}

\rev{One notable concern raised by participants was the potential censorship of information on abortion and contraception by GenAI systems across different states in the U.S. This is particularly important because major GenAI providers (e.g., OpenAI) typically moderate certain types of content, including hate speech, violence, self-harm, sexually explicit material, and illegal advice \cite{OpenAITransparency, OpenAIUsagePolicies}. Although most providers distinguish \textit{sensitive} from \textit{harmful} content (e.g., illegal activities), uncertainty remains about how criminalizing SRH topics like abortion are classified under these policies. Furthermore, moderation policies differ internationally based on government rules and Internet regulations. Examples include Ernie Bot (by Baidu in China), which reportedly refuses to answer politically sensitive questions \cite{mcdonell2023elusive}, and DeepSeek, which has been accused of censorship in accordance with Chinese government “public opinion guidance” regulations \cite{Guardian2025DeepSeek}.}

\rev{Finally, as discussed in \citet{friend2025abortion}, recent changes to social media content moderation policies—such as those by Meta, shifting from company-led monitoring to community-driven monitoring—may increase bias and inaccuracies in GenAI chatbots \cite{hendrix2025transcript}. Given that Meta trains its GenAI chatbot using data from public Facebook and Instagram posts, it is important to consider how polarized opinions on abortion expressed on social media may influence GenAI outputs \cite{CNN2024SocialMediaAI, MetaPrivacy}.}


\rev{\subsection{The Black Box Problem: Unclear Data Flows and Practices of GenAI Chatbots (RQ2)}} \label{DRQ2}

\rev{Although all participants expressed uncertainty about the data practices of GenAI companies, they revealed important aspects of their perceptions. Some participants incorrectly assumed that only their conversations (or only their prompts) were collected, without any personal or technical data. There was also widespread uncertainty about how responses were generated, with participants often assuming that the system rephrased online sources or searched databases for relevant information. While these assumptions may seem reasonable, since GenAI outputs, particularly from LLMs, often resemble summarized information, the underlying mechanism is different \cite{liu2025understanding}. It is true that the model’s knowledge often originates from data on the Internet, and if the system is explicitly integrated with a search tool (e.g., ChatGPT browsing), retrieval does occur \cite{oeding2025chatgpt}. However, rather than rephrasing documents or cross-referencing sources, the model generates text by predicting likely words based on patterns learned during training \cite{shanmugarasa2025sok, liu2025understanding}.}

\rev{Our findings exemplify the “black box problem,” a term used to describe users' (and creators') inability to understand and explain how GenAI systems work \cite{wadden2022defining, csahin2025unlocking, zednik2021solving}. This ambiguity introduces numerous vulnerabilities, including security risks, privacy violations, and bias, which become especially significant when GenAI is used for health or other sensitive purposes \cite{wadden2022defining}. Furthermore, participants' internal misperceptions, combined with a lack of awareness of available privacy protections, might distort how they noticed and interpreted data-related stimuli when interacting with GenAI chatbots, which in turn could affect privacy decision-making \cite{coopamootoo2014mental}.}

\rev{\subsection{Perceived Privacy Risks: From Oversharing to Overexposure (RQ3)}}

\subsubsection{\rev{Data Collection} in GenAI Chatbots vs. Other Tools for Seeking SRH Information}

Participants often perceived GenAI chatbots as carrying greater privacy risks than search engines, period-tracking apps, social media, or healthcare providers. This perception was primarily due to the distinct data collection methods of GenAI chatbots, which gather large amounts of detailed information in a conversational format. Our findings support this view by showing that participants shared sensitive SRH details in their GenAI prompts, providing detailed insight into participants' self-disclosure behaviors when using these tools.

 
In line with participants' beliefs, research indicates that GenAI tools collect large volumes of data through user interactions \cite{shanmugarasa2025sok}. For example, while period-tracking apps also handle sensitive data, their data collection is generally categorical and more limited \cite{tylstedt2023reimagining}. Similarly, search engines and social media pose privacy risks primarily through search history, location tracking, and browsing activity \cite{biega2014probabilistic, li2014all, degeling2018we}. Although healthcare providers and FDA-approved medical devices collect substantial volumes of sensitive personal data, these are rigorously governed by privacy regulations such as HIPAA and the General Data Protection Regulation (GDPR) \cite{drolet2017electronic}.

Why does seeking SRH information through GenAI chatbots lead users to share more personal data than when using other tools? Our findings suggest three key factors. First, \textit{anthropomorphism} \rev{\cite{li2022anthropomorphism}}, the attribution of human-like characteristics, such as being “nonjudgmental” or “compassionate,” to GenAI chatbots, has been associated with fostering user trust and engagement \cite{chen2021anthropomorphism, li2022anthropomorphism}. The perceived anthropomorphic nature of GenAI chatbots can directly predispose users to share more personal details by creating a false sense of intimacy and, consequently, safety. Second, \textit{interactivity} plays a critical role. As described by participants, GenAI chatbots “proactively” stimulate users to share personal information by asking follow-up questions, creating the sense of a dialogue with an attentive partner. This is particularly concerning because sensitive disclosures are not only insufficiently safeguarded, but also actively encouraged. Finally, the desire for \textit{personalized} responses—a common motivation for using GenAI chatbots—led participants to provide extensive personal details. This aligns with Privacy Calculus Theory, which posits that users continuously weigh perceived benefits against perceived risks, as observed with other digital tools \cite{meng2025talk}. However, unlike entering menstrual cycle dates into a period-tracking app or using a search engine, the conversational, open-ended design of GenAI prompts encourages richer disclosures. While social media platforms offer similar conversational interactions, users are typically aware that their posts are public or shared with a defined audience \cite{fiesler2017or}. In contrast, participants using GenAI chatbots were often unaware of—or held misconceptions about—data practices, raising significant concerns about transparency.

\subsubsection{Concerns About Intimate Surveillance and Criminalization}

Although participants generally felt comfortable seeking SRH information, most, particularly those living in restrictive states, expressed significant concerns regarding abortion and, in a few cases, SOGI-related topics. Following the recent \textit{overturn of Roe v. Wade} in the U.S., some participants feared that their abortion-related conversations could be subpoenaed by the government or flagged by the chatbot, potentially leading to criminalization, stigma, or harassment. Consequently, they lacked confidence in existing privacy protection laws, and often avoided such discussions.

This finding can be interpreted through the lens of Protection Motivation Theory (PMT) \cite{crossler2010protection}. According to PMT, individuals are more likely to adopt protective behaviors if they perceive the potential harm as severe (e.g., criminalization) and plausible (e.g., publicized cases involving digital SRH data), consider themselves vulnerable (e.g., residing in a restrictive state), and believe that the protective action (e.g., complete avoidance) is effective and feasible. In contrast, when perceived harms are abstract (e.g., selling anonymized SRH data to unknown recipients) or less immediate, participants tended to dismiss privacy threats. Moreover, SRH information types (e.g., STIs, abortion, endometriosis), associated with stronger negative emotional responses (e.g., fear, shame, stigma), elicited heightened privacy concerns, consistent with the Risk-as-Feelings Hypothesis \cite{loewenstein2001risk}.

Only two participants, both with technical backgrounds, expressed concern about sharing personal SRH details with GenAI chatbots regardless of the subject. While this caution may reflect greater digital literacy \cite{buchi2017caring}, this explanation remains speculative, as prior work has shown that even technically skilled individuals can make privacy errors, for example, IT employees unintentionally leaking company data through ChatGPT in 2022 \cite{petkauskas2023lessons}.

To what extent are concerns about surveillance and criminalization justified? Although there is no evidence that popular GenAI chatbots such as ChatGPT engage in proactive surveillance, law enforcement can subpoena companies to hand over user data, and companies are legally obligated to comply \cite{openai2025rowprivacy}. Additionally, participants’ concerns regarding privacy laws are well-founded, as sensitive health data within general-purpose GenAI chatbots is currently not protected by medical privacy regulations such as HIPAA and remains in a regulatory grey area \cite{minssen2023challenges}.

\subsubsection{\rev{Concerns About Targeted Advertising}}

\rev{Participants expressed concern that GenAI providers could use their data for commercial purposes, such as targeted advertising related to SRH products and services.} \rev{This concern is particularly salient because some GenAI companies have already indicated intentions to leverage user data for advertising. Recently, MetaAI announced that it will collect interactions with its AI tools—including both text and voice conversations—to serve targeted ads on its social media platforms, excluding users in the UK, EU, and South Korea \cite{OFlaherty2025MetaAIAds, Meta2025AIRecommendations}. The company stated that conversations on sensitive topics such as “sexual orientation,” “political views,” and “health” will not be used for advertising; however, questions remain regarding how these topics will be moderated \cite{Meta2025AIRecommendations}. Similarly, Google has announced plans to introduce AI-driven ads in its AI Mode feature within the search engine \cite{Google2025SearchAIBrandDiscovery}, Microsoft experimented with ads on Copilot a few years ago \cite{Mehdi2023BingPublishers}, and OpenAI has briefly considered introducing ads to ChatGPT \cite{Lomas2024ChatGPTAds}.} \rev{Given the rapid surge in research highlighting privacy risks of GenAI chatbots \cite{shanmugarasa2025sok}, the large volume of sensitive self-disclosure observed in our study and others \cite{kwesi2025exploring, yu2025exploring}, and repeated calls from cybersecurity experts, researchers, policymakers, and users for stronger privacy protections, these announcements by GenAI providers appear paradoxical and directly conflict with user expectations.}


\subsection{\rev{Insufficient Protective Measures for Safe GenAI Use (RQ4)}}

\subsubsection{Data Minimization: Primary Protective Measure}

Participants generally lacked sufficient knowledge to effectively mitigate privacy risks and rarely engaged with GenAI-specific privacy settings. Most often, they relied on withholding personal information, such as health details, PII, or location data.

Many GenAI providers (e.g., OpenAI) \rev{state in their privacy policies\footnote{All cited privacy policies refer to the 2025 versions available at the time of conducting the study.}} that they collect information about users’ IP addresses, country, and time zone, which can reveal location even if users do not explicitly share it \cite{OpenAIPrivacyPolicy}. Similarly, while participants minimized PII disclosures, GenAI platforms also collect account details, including names, contact information, and device identifiers, which can be used to trace data back to individual users \cite{OpenAIPrivacyPolicy}. According to ChatGPT’s privacy policy, these data types may be shared with third parties, including vendors, service providers, affiliates, government authorities, or industry peers, to comply with legal obligations. Although GenAI platforms typically aggregate or de-identify personal information, these data can still be re-identified when required by law \cite{OpenAIPrivacyPolicy}. Moreover, even partial or inaccurate data can enable GenAI chatbots to infer potentially identifying information about users \cite{staab2023beyond}. Therefore, users’ efforts to protect their privacy through data minimization may not fully align with their expectations.

\subsubsection{Challenges of Data Deletion in GenAI Systems}

While all participants viewed privacy concerns as a strong motivator for deleting their data, there were widespread misconceptions about how data deletion in GenAI chatbots actually works—for example, some assumed that sending a prompt like “Delete” would remove their data. Participants also expressed a lack of confidence in companies’ commitment to fully honoring deletion requests, suspecting that backups might still be retained. These concerns are valid, as deleting conversations typically removes them from a user’s account, but companies may retain internal copies. For instance, OpenAI’s privacy policy mentions temporary retention of logs without specifying the timeframe, and permanent deletion generally requires full account deletion. The challenge is further compounded because once user data is incorporated into model training, it is nearly impossible to remove without retraining the model; which is expensive and impractical \cite{ginart2019making}; opting into model training effectively precludes permanent deletion. In the SRH context, this is particularly concerning, as sensitive data embedded in models cannot be removed even if legally requested, and only a few participants engaged with model training settings, highlighting the need for greater attention to this issue.
\out{We advocate for stricter regulations requiring GenAI developers to set ``opt-out from model training'' as the default, ensuring users’ right to data deletion.}

\subsection{\rev{Recommendations}}

\rev{Here, we present practical and policy recommendations informed by our findings. Table \ref{tab:codes-description} provides a summary.}

\subsubsection{\rev{Public awareness and involvement}}
\rev{We emphasize the need to improve public GenAI literacy. We advocate for robust educational campaigns on responsible use, targeting both patients and clinicians, to foster effective clinician-patient-GenAI \textit{collaboration} \cite{kim2024adaptive}. Clinicians and patients should engage in open, transparent discussions about the benefits and risks of using GenAI chatbots for SRH management, ensuring proper verification practices, appropriate levels of trust, and safe disclosure.}

\moved{Given the limited awareness and use of data protection strategies among our participants, we recommend greater user involvement in designing and developing \rev{\textit{human-centered}} privacy protections in GenAI chatbots. This is particularly important for vulnerable users seeking sensitive health information \rev{(e.g., people assigned female at birth or from diverse SOGI subgroups)}, who may face risks such as criminalization, harassment, or stigmatization \rev{\cite{meister2022digital, mcdonald2022privacy}}. To align protections with users’ needs, we suggest organizing co-design workshops within a Participatory Action Research (PAR) framework \cite{livingston2018participatory}, where end-users collaborate with designers, developers, and researchers to explore privacy mechanisms such as real-time reminders, opt-out options, and other safeguards for sensitive data collection.}

\subsubsection{\rev{Privacy by Design}}
\rev{As individuals increasingly use GenAI chatbots for deeply personal and intimate queries, developers should design systems with self-disclosure considerations in mind. GenAI providers must adopt a Privacy by Design approach (e.g., \cite{al2025framework}), integrating data protection into the core design and architecture of GenAI systems. For example, companies should invest in and transparently communicate technical and design approaches to data deletion, including: (a) advancing machine-unlearning techniques and auditing tools to ensure stronger guarantees for removing specific data from models over time \cite{sun2025unlearning, liu2024revisiting}, (b) implementing conservative filtering and data-minimization strategies to remove or obfuscate sensitive content before user data enters the training pipeline \cite{brown2022does, pal2024empirical},} \moved{and (c) setting ``opt-out from model training'' as the default, thereby safeguarding users’ rights to data deletion or clearly communicating alternative options \rev{\cite{keller2023defining, bui2022opt}}. Additionally, companies could employ Reinforcement Learning from Human Feedback (RLHF) \cite{ouyang2022training} to align chatbot responses with ethical standards, train models to handle sensitive prompts responsibly, and redirect users to trusted health resources.}

\subsubsection{\rev{Interactive privacy}}
\rev{While GenAI interactivity has been identified as both a facilitator of self-disclosure and a potential privacy limitation, this feature can become an asset if applied thoughtfully. Since GenAI can recognize sensitive discussions, it can leverage dynamic, interactive strategies to support users in making \textit{informed} decisions.} \moved{For instance, it can request explicit consent for data collection in real time, provide gentle discouragement from oversharing personal information, issue safety warnings (e.g., legal consequences), offer privacy nudges and reminders about available privacy settings \rev{\cite{acquisti2017nudges, murmann2021design}}, and redirect users to trusted resources. Moreover, providers should clearly identify GenAI chatbots as AI-based tools—not human or confidential healthcare professionals (e.g., via generated outputs)—to prevent unrealistic user expectations.}

\subsubsection{\rev{Improved transparency}}

\rev{Although the “black box problem,” discussed in \S\ref{DRQ2}, is well-known and has been explored in prior research \cite{zednik2021solving, haque2023explainable}, it is especially concerning in the context of sensitive conversations. Providers should adopt more transparent and explainable AI (XAI) approaches \cite{zednik2021solving, haque2023explainable, Kosinski2024BlackBoxAI}, such as open-source models, data visualizations, user-friendly transparency dashboards, and alerts for high-risk use cases where opaque processing could have serious consequences. These measures help users better understand how AI processes data and generates outputs, fostering greater awareness of the system’s limitations.}

\subsubsection{\rev{GenAI regulations}}

\rev{GenAI is an emerging and complex technology, still in the early stages of developing clear ethics, governance, and regulation. Existing frameworks (e.g., the EU AI Act) have been criticized for their incompleteness \cite{wachter2023limitations}, challenges that are particularly acute in health-related contexts. Given the discrepancies between user expectations observed in this study and the data practices of GenAI providers, there is an urgent need for \textit{global} data protection regulations that ensure \textit{equitable} protections across international users. To begin addressing these issues, we recommend that GenAI providers introduce} \moved{a dedicated ``Health'' model (similar to custom ``GPTs'' in ChatGPT) designed to comply with medical privacy laws, incorporating encryption and secure data handling to support safe health discussions while maintaining accessibility.} \rev{Additionally, healthcare providers could consider deploying GenAI models under strict local data controls to enable secure use among patients.}

\rev{We recognize that such regulations may slow innovation in GenAI; however, rapid advancement should not come at the cost of safe and ethical implementation. To mitigate potential delays, we advocate for \textit{collaborative} policymaking between government and industry (co-regulation \cite{latzer2013self, cantero2024artificial}) rather than relying solely on government or self-regulation. Co-regulation can support initiatives such as regulatory sandboxes \cite{yordanova2024regulating}, which provide controlled environments for testing new rules.}

\subsection{Comparison with Prior Work}

Our study builds on prior research in two key areas: (1) users' privacy and safety perceptions and concerns regarding SRH data, expanding previous findings beyond FemTech to include general-purpose GenAI chatbots; and (2) users' perceptions of and experiences with GenAI chatbots in the sensitive context of SRH information seeking. In this subsection, we outline how these differences led to contrasts with other studies.

Consistent with prior findings, GenAI tool adoption is shaped by utility, usability, credibility, affordability, and anthropomorphism; however, our participants emphasized additional factors specific to SRH \cite{al2023investigating, alanezi2024factors, zhang2024s}. They highlighted the importance of equitable access to neutral, comprehensive information and raised concerns about location-based barriers and censorship, particularly in conservative states.

\rev{A recent study \cite{wang2025mental} explored participants' mental models of GenAI chatbots in a different context, focusing on task-based interactions (i.e., booking a hotel). Yet, we observed some conceptual overlap. For instance, \citet{wang2025mental} found that users believed the chatbot directly shared data with its provider, mirroring our participants’ belief that data were sent to the provider for service improvement. Moreover, \citet{wang2025mental} also found that users’ perceptions of the chatbot's parent company shaped trust. We observed a similar pattern, although some of our participants held incorrect beliefs about the parent company (e.g., believing ChatGPT was owned by Google), which could lead to misplaced trust.}

\rev{In another} interview study involving participants seeking mental health information in GenAI chatbots \cite{kwesi2025exploring}, \rev{participants were primarily concerned about sharing data with employers and insurers, resulting in employability harms. In contrast to mental health topics,} SRH-related conversations raised concerns primarily about government surveillance and criminalization. Our participants also identified GenAI-specific risks beyond general data breaches, comparing them to privacy risks in other tools. Kwesi et al. \cite{kwesi2025exploring} also noted that while emotional and psychological data is deeply personal, \rev{participants did not perceive} obvious real-world exploit pathways that would prompt more cautious behavior. Findings related to \rev{our participants' perceptions of} SRH information seeking align with this observation, except for abortion or other potentially criminalized SRH topics. \rev{In these cases,} fears were likely amplified by real-world instances in which some participants might face prosecution based on their digital SRH data. Another key contrast with \cite{kwesi2025exploring} is that, while participants in their study believed that their conversations were HIPAA-protected, our participants lacked confidence in privacy protections, especially in the context of conflicting abortion regulations.

Compared to FemTech research, our findings related to government surveillance and criminalization align with users’ risk perceptions of other FemTech tools \cite{dewan2024teen, mcdonald2023did, mehrnezhad2023my, cao2024deleted}. However, participants in our study expressed additional concerns about GenAI chatbots collecting more data than other FemTech tools (e.g., period-tracking apps) and potentially \textit{recognizing} illegal or sensitive topics, a perception likely influenced by the chatbots' attributed intelligence and anthropomorphism. Furthermore, among participants who used risk mitigation strategies, most protected themselves by minimizing or attempting to delete data, echoing findings from prior studies with users of both GenAI \cite{zhang2024s} and FemTech \cite{mehrnezhad2023my, cao2024deleted, mcdonald2023did}. Only a few participants configured GenAI-specific privacy settings, such as opting out of model training.

\rev{Future research should employ data-donation studies to examine how people interact with GenAI chatbots during sensitive SRH conversations; for example, how they formulate prompts, what they disclose, how the chatbot responds, and how follow-up questions unfold. Measurement studies could also test whether chatbot responses vary by location (e.g., via VPN), reflecting participants’ concerns about potential information censorship. Additionally, although our qualitative data showed no clear differences in protections between participants in restrictive and non-restrictive states, aside from avoiding abortion-related queries due to criminalization concerns, large-scale quantitative studies may reveal distinct patterns between these groups.}

\section{Conclusion}
This study examines users’ experiences with GenAI chatbots for seeking SRH information. The findings indicate that such chatbots are increasingly used due to their perceived utility, usability, credibility, accessibility, and capacity to provide empathetic, non-judgmental support. At the same time, participants viewed GenAI chatbots as posing heightened privacy risks compared to other information sources, citing concerns about the large volume of personal data collected, the use of data for model training, potential government access or surveillance, user profiling, data commercialization, and lack of regulatory protections. Although participants were generally willing to accept these risks in exchange for perceived benefits, they expressed reluctance to seek abortion-related information or advice, particularly in abortion-restrictive states. Moreover, most participants did not effectively adopt GenAI-specific strategies to mitigate these risks. To our knowledge, this is the first study to investigate users’ privacy and safety experiences with GenAI chatbots in the context of SRH information seeking. These findings offer important implications for the design of more user-centered, privacy-preserving, and ethically grounded GenAI technologies.

\begin{acks}
This research was supported by INCIBE’s strategic SPRINT (Seguridad y Privacidad en Sistemas con Inteligencia Artificial) project C063/23, funded by the EU NextGenerationEU program through the Spanish Government’s Plan de Recuperación, Transformación y Resiliencia; by the Spanish Government under grant PID2023-151536OB-I00; by the Generalitat Valenciana under grant CIPROM/2023/23; and by the UK Engineering and Physical Sciences Research Council (EPSRC) under award RE16677. We are grateful to all participants for generously sharing their time and experiences. We also acknowledge the UKRI Centre for Doctoral Training in Safe and Trusted Artificial Intelligence for providing academic training (e.g., seminars and masterclasses) that contributed to this research.
\end{acks}

\bibliographystyle{ACM-Reference-Format}
\bibliography{sample-base}

\appendix


\section{Study Materials}

\subsection{Screening Survey}

\label{Survey}

\paragraph{[\textit{Use of GenAI}]}

In this section, we will ask you questions about your use of \textbf{generative AI tools}. These are programs that create new content, such as text, images, video, or other data. Examples include \textbf{OpenAI's ChatGPT, Microsoft Copilot, Google’s Gemini,} and \textbf{Meta's Llama 2}. Please note that in this section, we are \textbf{NOT} referring to voice assistants such as Amazon Alexa, Apple Siri, Samsung Bixby, etc.

1. GenAI usage -- Have you ever used a \textbf{generative AI tool}?

- Yes, I have used a generative AI tool (1)  
- No, I have not used a generative AI tool (2)  

2. How often do you use a \textbf{generative AI tool}?  

- Daily (1)  
- Weekly (2)  
- Monthly (3)  
- Other (please specify) (4)  
- Prefer not to say (5)  

3. Have you ever used a generative AI tool to \textbf{seek information or advice about sexual and reproductive health}?  

- Yes, I have used a generative AI tool for this purpose (1)  
- No, I have not used a generative AI tool for this purpose (2)  

4. Please specify the generative AI tool(s) you have used for seeking sexual and reproductive health information. (Check all that apply)  

- OpenAI's ChatGPT (1)  
- Google's Gemini (2)  
- Microsoft Copilot (3)  
- Meta's Llama 2 (4)  
- Other (please specify) (5)

5. What specific sexual or reproductive health topic(s) did you seek information about using a generative AI tool? (Check all that apply)

- Contraception methods (1)  
- Pregnancy-related information (2)  
- Sexual health and wellness (3)  
- Menstrual health and fertility (4)  
- Other (please specify) (5)  
- Prefer not to say (6)  %

6. Please use the slider below to indicate your level of concern about how \textbf{generative AI tools} handle your data, with 0 meaning “not at all concerned” and 10 meaning “extremely concerned.”  

\paragraph{[\textit{Use of Voice Assistants}]}  

In this section, we will ask you questions about your use of \textbf{voice assistants}. Voice assistants are devices or applications that can understand human language and respond to questions, provide information, or assist with tasks using speech. They can be found on mobile phones, smart speakers, smartwatches, and other devices. Some well-known examples include \textbf{Amazon Alexa, Google Assistant, Apple’s Siri, Microsoft Cortana, Samsung Bixby,} and \textbf{Blackberry Assistant}. Please note that in this section, we are \textbf{NOT} referring to generative artificial intelligence (AI) tools such as ChatGPT, Microsoft Bing, or similar systems.  

7. Have you ever used a \textbf{voice assistant}?  

- Yes, I have used a voice assistant (1)  
- No, I have not used a voice assistant (2)  

8. How frequently do you use a \textbf{voice assistant}?  

- Daily (1)  
- Weekly (2)  
- Monthly (3)  
- Other (please specify) (4)  
- Prefer not to say (5)  

9. Have you ever used a \textbf{voice assistant} to seek information or advice about \textbf{sexual and reproductive health}?  

- Yes, I have used a voice assistant for this purpose (1)  
- No, I have not used a voice assistant for this purpose (2)  

10. Please specify which \textbf{voice assistant(s)} you have used to seek information about \textbf{sexual and reproductive health}. (Check all that apply.)  

- Amazon Alexa (1)  
- Google Assistant (2)  
- Apple's Siri (3)  
- Microsoft Cortana (4)  
- Samsung Bixby (5)  
- Blackberry Assistant (6)  
- Other (please specify) (7)  

11. What specific \textbf{sexual and reproductive health topic(s)} have you sought information about using a \textbf{voice assistant}? (Check all that apply.)  

- Contraception methods (1)  
- Pregnancy-related information (2)  
- Sexual health and wellness (3)  
- Menstrual health and fertility (4)  
- Other (please specify) (5)  
- Prefer not to say (6)  

12. Please use the slider below to indicate your level of concern about how \textbf{voice assistants} handle your data, with 0 meaning “not at all concerned” and 10 meaning “extremely concerned.”  

\paragraph{[\textit{Abortion Attitudes}]}  

13. Please indicate the extent to which you agree or disagree with the following statement: \textbf{"Abortion is health care."}  

- Strongly Disagree (1)  
- Disagree (2)  
- Neither Agree nor Disagree (3)  
- Agree (4)  
- Strongly Agree (5)  

\paragraph{[\textit{Technical Background}]}  

14. Do you have education or work experience in any \textbf{information technology (IT) fields}, such as \textbf{Computer Science, Software Engineering, or App Development}?  

- Yes (1)  
- No (2)  

15. Do you regularly use any of the following methods to protect your privacy?  

- Virtual Private Networks (VPNs) (1)  
- End-to-end encryption (e.g., encrypted emails) (2)  
- Private browsing (e.g., Incognito mode) (3)  
- Tor Browser (4)  
- DuckDuckGo search engine (5)  
- Password managers (6)  
- Other (please specify) (7)  
- None of the above (8)  

\paragraph{[\textit{Demographics}]}  

16. Next, we will ask several questions about \textbf{diversity and inclusion}. Your participation in this study will not be affected by your answers. Please select the group that best describes you from the options below.  

- White or Caucasian (1)  
- Black or African American (2)  
- Asian or Asian American (3)  
- Native Hawaiian or Other Pacific Islander (4)  
- American Indian or Alaska Native (5)  
- Arab or North African (6)  
- Mixed or multiple groups (please specify) (7)  
- Prefer to self-describe (8)  
- Prefer not to say (9)  

17. What is your gender?  

- Woman (1)  
- Man (2)  
- Non-binary or prefer to self-describe (3)  
- Prefer not to say (4)  

18. What is the highest level of education you have completed?  

- Less than high school (1)  
- High school/GED or equivalent (2)  
- Some college or university but no degree (3)  
- Associate degree (4)  
- Bachelor's degree or equivalent (5)  
- Master's degree or equivalent (6)  
- Doctoral or professional degree (PhD, JD, MD) (7)  
- Other, please specify (8)  
- Prefer not to say (9)  

19. How would you describe your current employment status?  

- Unemployed (1)  
- Employed full-time (2)  
- Employed part-time or casually (3)  
- Self-employed (4)  
- Student (5)  
- Retired (6)  
- Other (please specify) (7)  
- Prefer not to say (8)  

20. Please select the option that best describes your total annual household income after taxes.  

- Less than \$10,000 (1)  
- \$10,000–\$15,999 (2)  
- \$16,000–\$19,999 (3)  
- \$20,000–\$29,999 (4)  
- \$30,000–\$39,999 (5)  
- \$40,000–\$49,999 (6)  
- \$50,000–\$59,999 (7)  
- \$60,000–\$69,999 (8)  
- \$70,000–\$79,999 (9)  
- \$80,000–\$89,999 (10)  
- \$90,000–\$99,999 (11)  
- \$100,000–\$149,999 (12)  
- More than \$150,000 (13)  
- Prefer not to say (14)  

\paragraph{[\textit{Final Question}]}  
Please briefly explain why you are interested in participating in this study.

\subsection{Interview Script}
\label{InterviewScript}

\subsubsection{General Views and Experiences}

In this interview, I will focus on your use of \textbf{generative AI tools} specifically for seeking information or advice on any aspect of sexual and reproductive health (SRH), including menstrual health, fertility, sexual health, pregnancy, contraception, and abortion. To clarify, generative AI tools are technologies that generate new data; examples include ChatGPT, Microsoft Copilot, Google Gemini, and others. In the screener, you reported that you have used [tools], is that correct?

[\textit{If multiple tools have been used}]: Which of these tools have you used the most for seeking information about SRH, and why? I will focus my questions on [tool] specifically; however, please feel free to share your experiences with [other tools] whenever you think it is relevant or important.

1. Can you describe a situation in which you used [tool] to seek information or advice related to SRH? 

2. What motivated you to use [tool] specifically for seeking SRH information or advice?  

   - Why did you choose [tool] instead of other generative AI tools?  
   - [\textit{If the tool is ChatGPT}]: Did you use a custom version of ChatGPT or the general version?

3. Have you noticed any benefits of using [tool] for seeking SRH information?

4. Have you noticed any drawbacks or challenges when using [tool] for seeking SRH information?

5. What other sources of SRH information do you currently use or have used in the past?

6. Did you use [tool] as your primary source of information on SRH?

7. How do you think [tool] differs from other sources of information for seeking SRH information?

8. Are there situations or circumstances in which you prefer using [tool] over other sources of SRH information?

9. In contrast, are there situations or circumstances in which you prefer using other sources of information over [tool] for SRH queries?

\subsubsection{Interaction with GenAI tools}

10. Can you describe the process you followed when interacting with [tool] to seek SRH information?  

    - What prompts did you use when interacting with [tool] to seek SRH information?  
    - What information, if any, did you share with [tool] when seeking SRH information (for example, personal information or health- and non-health-related data)?

11. How effective was [tool] in helping you accomplish your goal, and why?

12. How did interacting with [tool] compare to interacting with a human, such as a friend or healthcare professional, when seeking SRH information or advice?

I will now ask several questions about the information generated by [tool] and how you perceive its quality. I will clarify any terms before each question.

14. How complete do you think the information on SRH provided by [tool] was, and why?  
[Provide clarification: information is complete when all relevant elements and details are present and no essential information is missing]

15. How consistent do you think the information on SRH provided by [tool] was compared to other sources, and why?  
[Provide clarification: consistency means that the information given by [tool] does not contradict information from other sources]

16. How authentic do you think the information on SRH provided by [tool] was, and why?  
[Provide clarification: information is authentic when it has an origin supported by unquestionable evidence and is verified]  

- Where do you think the SRH information provided by [tool] came from?

17. How authoritative do you think [tool] is as a source of information on SRH, and why?  
[Provide clarification: authoritative sources are widely recognized or generated by experts in a specific field]

18. Did you verify the SRH information provided by [tool], and why or why not?

19. How did you use or act on the SRH information you received from [tool]?  

- Did you make any health-related decisions based on the information you received from [tool]?

20. In your opinion, are there any potential safety risks or harms posed by using [tool] to seek SRH information? If so, please describe them.

21. How do you feel about these safety risks?

22. How do you think the potential safety risks compare to, or differ between, [tool] and other sources of SRH information, such as web-based sources or healthcare professionals?

23. What potential consequences or harms to users do you think may result from these safety risks?

24. Do you take any active steps to protect yourself against the safety risks of seeking SRH information using [tool]? If so, please describe these steps.

\subsubsection{Beliefs About Data Practices}

I will now ask you questions about your understanding of the data collected by [tool] and how it flows within [tool]’s system. By “data flow,” I mean how the information collected by [tool] moves from one place to another and who might receive it. There are no right or wrong answers. Please try to use your imagination and think aloud. Don’t worry about specific terminology.

25. What types of data do you think [tool] collected when you used it to seek SRH information?  

- Are both the prompts you enter and the responses generated by [tool] collected?  
- What other data, if any, do you think [tool] has collected about you in general, for example, during registration or while using [tool]? What data, if any, do you think [tool] collects automatically on its own?  
- Where do you think this data is stored, and for how long?  
- Are there any measures you believe were taken to protect this data?

26. How do you feel about the collection of your conversation data?  

- What do you think the purpose or purposes of [tool] collecting conversation data are?  
- Who do you think has access to your conversation data, if anyone, and for what purposes?

27. How do you feel about the collection of your account data?  

- What do you think the purpose or purposes of [tool] collecting this data are?  
- Who do you think has access to the data collected by [tool], if anyone, and for what purposes?

28. What do you believe happens to the data you provide to [tool] when seeking SRH information, before it generates a response? How do you think this data is processed by [tool] to generate a response?  

29. What do you think [tool] learned or inferred about you or others based on your SRH-related interactions?  

- How do you feel about [tool] learning or inferring this?  
- How confident do you feel that the information learned or inferred is accurate?  
- What do you think the purpose or purposes of [tool] learning and inferring this data about you or others are?

\subsubsection{Data Deletion}

30. Have you ever deleted any of the data collected or used by [tool]?  

- [\textit{If yes}] What data did you delete?

31. Why did you delete your data?  

32. How did you delete your data, and can you describe the process?  

33. What do you think happened to your data after you deleted it?  

34. Did you export a copy of your data before deleting it, and why or why not?  

[\textit{Hypothetical: If participants did not delete data collected or used by [tool], ask the following questions:}]

35. What would motivate you to delete data collected or used by [tool]?  

- What type of data would you choose to delete?

36. How would you delete each of these data types, and can you describe the process?  

37. What do you think would happen to your data after you deleted it?  

38. Would you export a copy of your data before deleting it, and why or why not?

\subsubsection{Privacy Risks and Mitigation Strategies}

Privacy refers to an individual’s right to control their personal information, including how it is collected, used, shared, and who can access it. Data security refers to the protection of data from potential threats.

39. In your opinion, are there any potential privacy or security risks when using [tool] or other generative AI tools to seek sexual and reproductive health information?  

40. Who do you think would benefit from any privacy or security risks, if at all?  

41. How do you feel about the privacy and security risks posed by seeking SRH information in [tool]?  

42. Have you ever wanted to ask or share something with [tool] but chose not to because of privacy or security concerns?  

43. How do you think the potential privacy and security risks of seeking SRH information compare to, or differ between, [tool] and other sources, such as web-based platforms or healthcare providers?  

- Are there any privacy or security risks that are specific to [tool] and generative AI chatbots in general?

44. What potential consequences or harms to users do you think may result from the privacy and security risks of using [tool] to seek SRH information?  

45. Do you take any active steps to protect yourself against the privacy and security risks of seeking SRH information in [tool]? If so, please describe these steps.  

46. Have you ever interacted with the privacy or security settings of [tool]?  

- Are you aware of, or have you used, [\textit{each available privacy or security setting in the GenAI chatbot used by the participant}]?

47. Are there any other potential risks, besides privacy, security, and safety, that you think might arise from using [tool] to seek SRH information that I haven’t asked you about?  

48. Overall, do you think [tool] is worth using for this purpose, considering the potential risks you described? Why or why not?

\subsubsection{Privacy Legislation and Abortion}

49. Are you aware of any privacy laws in your country or [state] that protect people’s digital data?  

- Who do you think this law or these laws are designed to protect?  
- What protections does this law, or do these laws, offer?  
- Does this law, or do these laws, protect people’s personal information or digital data specifically related to SRH when using [tool] or other generative AI tools?

50. According to your Prolific/screener data, you live in [state]. Can you describe the current legal status of abortion there?  

- [\textit{If the person is pro-choice-oriented}]: Would you use [tool] to seek information or advice about abortion care access, and why or why not?  
- [\textit{If the person is not pro-choice-oriented}]: What are your thoughts on using generative AI tools, such as [tool], to seek information or advice about abortion care access?

51. Can you describe any specific ways you believe abortion-related conversations might be processed, stored, used, or shared by [tool]?

52. To what extent do you feel confident that [privacy law] adequately protects people’s abortion-related conversations in [tool] or other generative AI tools, and why or why not?

\subsubsection{Suggestions and Recommendations}

53. Based on our discussion, are there any specific improvements you would like the developers of [tool] or other generative AI tools to make to enhance your privacy, security, or safety when seeking SRH information?  

54. What other measures would you like governments and policymakers to take to address users’ concerns regarding [tool] and other generative AI tools?  

55. Would you like to share or discuss any additional thoughts?


\end{document}